\documentclass[12pt]{emulateapj}
\usepackage{color}
\usepackage{threeparttable}

\newcommand\ha{H$\alpha$~}

\shorttitle{SXDF-ALMA 1.5 arcmin$^2$ deep survey}
\shortauthors{Tadaki et al.}

\begin{document}


\title{SXDF-ALMA 1.5 arcmin$^2$ deep survey. A compact dusty star-forming galaxy at z=2.5.}


\author{Ken-ichi Tadaki\altaffilmark{1}, Kotaro Kohno\altaffilmark{2,3}, Tadayuki Kodama\altaffilmark{4,5}, 
Soh Ikarashi\altaffilmark{6},
Itziar Aretxaga\altaffilmark{7}, 
Stefano Berta\altaffilmark{1},
Karina I. Caputi\altaffilmark{6}, 
James S. Dunlop\altaffilmark{8}, 
Bunyo Hatsukade\altaffilmark{4}, 
Masao Hayashi\altaffilmark{4}, 
David H. Hughes\altaffilmark{7}, 
Rob Ivison\altaffilmark{8,9}, 
Takuma Izumi\altaffilmark{2}, 
Yusei Koyama\altaffilmark{4}, 
Dieter Lutz\altaffilmark{1},
Ryu Makiya\altaffilmark{2}, 
Yuichi Matsuda\altaffilmark{4,5},
Kouichiro Nakanishi\altaffilmark{4,5,10}, 
Wiphu Rujopakarn\altaffilmark{11,12}, 
Yoichi Tamura\altaffilmark{2},
Hideki Umehata\altaffilmark{2,9}, 
Wei-Hao Wang\altaffilmark{13,14}, 
Grant W. Wilson\altaffilmark{15},
Stijn Wuyts\altaffilmark{1}, 
Yuki Yamaguchi\altaffilmark{2},
Min S. Yun\altaffilmark{15}
%
}


\affil{\altaffilmark{1} Max-Planck-Institut f{\"u}r extraterrestrische Physik (MPE), Giessenbachstr., D-85748 Garching, Germany; tadaki@mpe.mpg.de}
\affil{\altaffilmark{2} Institute of Astronomy, The University of Tokyo, 2-21-1 Osawa, Mitaka, Tokyo 181-0015, Japan}
\affil{\altaffilmark{3} Research Center for the Early Universe, The University of Tokyo, 7-3-1 Hongo, Bunkyo, Tokyo 113-0033, Japan}
\affil{\altaffilmark{4} National Astronomical Observatory of Japan, 2-21-1 Osawa, Mitaka, Tokyo 181-8588, Japan}
\affil{\altaffilmark{5} Department of Astronomical Science, SOKENDAI (The Graduate University for Advanced Studies), Mitaka, Tokyo 181-8588, Japan}
\affil{\altaffilmark{6} Kapteyn Astronomical Institute, University of Groningen, P.O. Box 800, 9700AV Groningen, The Netherlands}
\affil{\altaffilmark{7} Instituto Nacional de Astrof{\'i}sica, {\'O}ptica y Electr{\'o}nica (INAOE), Luis Enrique Erro 1, Sta. Ma. Tonantzintla, Puebla, Mexico }
\affil{\altaffilmark{8} Institute for Astronomy, University of Edinburgh, Royal Observatory, Blackford Hill, Edinburgh EH9 3HJ, UK}
\affil{\altaffilmark{9} European Southern Observatory, Karl-Schwarzschild-Strasse 2, Garching bei Munchen, Germany}
\affil{\altaffilmark{10} Joint ALMA Observatory, Alonso de C{\'o}rdova 3107, Vitacura 763-0355, Santiago de Chile}
\affil{\altaffilmark{11} Department of Physics, Faculty of Science, Chulalongkorn University, 254 Phayathai Rd, Pathumwan, Bangkok 10330, Thailand}
\affil{\altaffilmark{12} Kavli Institute for the Physics and Mathematics of the Universe (WPI), Todai Institute for Advanced Study, University of Tokyo, 5-1-5 Kashiwanoha, Kashiwa, 277-8583, Japan}
\affil{\altaffilmark{13} Institute of Astronomy and Astrophysics, Academia Sinica, Taipei, Taiwan}
\affil{\altaffilmark{14} Canada-France-Hawaii Telescope, HI, USA}
\affil{\altaffilmark{15} Department of astronomy, University of Massachusetts, Amherst, MA 01003, USA}



\begin{abstract}
We present first results from the SXDF-ALMA 1.5 arcmin$^2$ deep survey at 1.1 mm using Atacama Large Millimeter Array (ALMA).
The map reaches a 1$\sigma$ depth of $55~\mu$Jy/beam and covers 12 H$\alpha$-selected star-forming galaxies at $z = 2.19$ or $z=2.53$.
We have detected continuum emission from three of our H$\alpha$-selected sample, including one compact star-forming galaxy with high stellar surface density, NB2315-07.
They are all red in the rest-frame optical and have stellar masses of log $(M_*/M_\odot)>10.9$ whereas the other blue, main-sequence galaxies with $\log(M_*/M_\odot)$=10.0-10.8 are exceedingly faint, $<290~\mu$Jy ($2\sigma$ upper limit).
We also find the 1.1 mm-brightest galaxy, NB2315-02, to be associated with a compact ($R_e=0.7\pm0.1$ kpc), dusty star-forming component.
Given high gas fraction (44$^{+20}_{-8}$\% or $37^{+25}_{-3}$\%) and high star formation rate surface density (126$^{+27}_{-30}~M_\odot$yr$^{-1}$kpc$^{-2}$),
the concentrated starburst can within less than $50^{+12}_{-11}$ Myr build up a stellar surface density matching that of massive compact galaxies at $z\sim2$, provided at least 19$\pm3$\% of the total gas is converted into stars in the galaxy center.
On the other hand, NB2315-07, which already has such a high stellar surface density core, shows a gas fraction (23$^{+8}_{-8}$\%) and is located in the lower envelope of the star formation main-sequence.
This compact less star-forming galaxy is likely to be in an intermediate phase between compact dusty star-forming and quiescent galaxies.

\end{abstract}


\keywords{galaxies: evolution --- galaxies: high-redshift --- galaxies: ISM}



\section{Introduction}
\label{sec;intro}

One of the fundamental questions in galaxy formation is when and how massive galaxies are formed at high redshift.
Massive quiescent galaxies (QGs) already existed at $z\geq2$ when the cosmic star formation activity peaked (e.g. \citealt{2009ApJ...700..221K}). 
They are extremely compact with an effective radius of only $R_\mathrm{e}\sim1$ kpc in the rest-frame optical, which is a factor of 4--5 smaller than the size of local QGs at similar stellar masses. (e.g., \citealt{2007MNRAS.382..109T, 2008ApJ...677L...5V, 2012ApJ...746..162N}). 
This discovery has challenged galaxy evolution theories and raised the major question of how the compact QGs subsequently grow in size.
It has become widely accepted that the observed size evolution is due to a combination of two evolutionary channels: \textit{slow} and \textit{fast} modes.
In the slow mode, new, larger quenched galaxies continuously add to quiescent population at $0<z<2$ (e.g., \citealt{2010ApJ...713..738W, 2015ApJ...799..206B}). 
In the fast mode, compact star-forming galaxies (SFGs) are quenched to become compact QGs at $z\sim2$ and subsequently puff up through repeated dry minor mergers (e.g., \citealt{2009ApJ...699L.178N, 2010ApJ...709.1018V}). 
While the size growth of quiescent galaxies has been partially understood, the formation mechanisms of compact stellar components, in the fast mode, remain poorly done.

Theoretical studies predict that a 1 kpc scale core can be produced by dissipative processes such as gas-rich major mergers or disk instability-driven inflow within galaxies \citep{2010ApJ...722.1666W, 2011ApJ...730....4B, 2014MNRAS.438.1870D,2015MNRAS.449..361W}.
Such violent processes should happen on a short timescale so that external gas accretion does not re-grow galaxy disks.
While typical massive SFGs have a large disk with $R_\mathrm{e}\sim4$ kpc at $z\sim2$ (e.g., \citealt{2014ApJ...788...28V}), recent high-resolution ALMA imaging reveals that submillimeter bright galaxies (SMGs) have a dusty star-forming core whose size is four times smaller than their stellar component (\citealt{2015ApJ...799...81S}, see also \citealt{2014arXiv1411.5038I}). 
Moreover, massive compact SFGs have been discovered at $z\sim2$ (e.g., \citealt{2011ApJ...742...96W, 2013ApJ...765..104B, 2014ApJ...780...77T}) and 
their high stellar surface densities suggest that they can directly evolve into compact QGs without structural transformation, simply by quenching star formation.
The number density evolution of compact SFGs supports the production of the observed number densities of compact QGs over the same epoch with a quenching timescale of 0.3--1.0 Gyr \citep{2013ApJ...765..104B}.
Deep near-infrared slit spectroscopy also suggests that their dynamical masses are comparable to their stellar masses, indicating a short quenching timescale \citep{2014ApJ...795..145B, 2014Natur.513..394N}.
All these results point towards an evolutionary scenario in the fast mode whereby extended SFGs evolve into compact SFGs through a violent starburst phase, and then become compact QGs on a short quenching timescale.

\begin{figure}
\begin{center}
\includegraphics[scale=0.93]{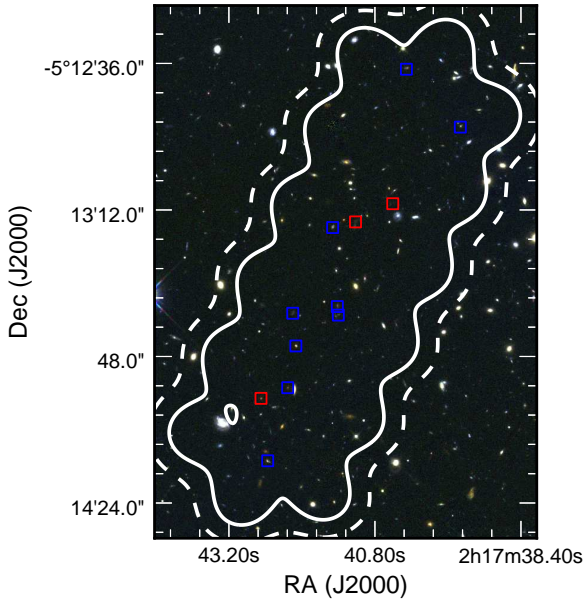}
\end{center}
\caption{
12 H$\alpha$-selected SFGs in the SXDF-ALMA 1.5 arcmin$^2$ deep field. Red and blue squares indicate 1.1 mm sources and non-detection, respectively. A white solid and dashed line show the areas where the primary beam correction is below 20\% and 50\%, respectively. 
}\label{fig:FoV}
\end{figure}

In this letter, we present 1.1 mm properties of SFGs at $z\sim2$, along with the discovery of a galaxy with an extremely compact, dusty star-forming component. 
These are the first results from the SXDF-ALMA 1.5 arcmin$^2$ deep survey, which is unconfused (FWHM=0\arcsec.5) and unbiased (i.e., blind) observations with Atacama Large Millimeter Array (ALMA).
The ALMA observations target a part of the SXDF-UDS-CANDELS field, where {\it Hubble Space Telescope (HST)} high resolution images are available \citep{2011ApJS..197...35G, 2011ApJS..197...36K}.
We only focus on 12 SFGs identified by a H$\alpha$ narrow-band imaging survey with Subaru Telescope \citep{2013ApJ...778..114T, 2013IAUS..295...74K}. %
Survey design and source catalog of the ALMA observations are described in detail in Kohno et al. in preparation.
We assume the Chabrier initial mass function (IMF; \citealt{2003PASP..115..763C}) and adopt cosmological parameters of $H_0$ =70 km s$^{-1}$ Mpc$^{-1}$, $\Omega_{\rm M}$=0.3, and $\Omega_\Lambda$ =0.7.

\section{Sample and Data}
\label{sec;data}

The SXDF-ALMA 1.5 arcmin$^2$ deep survey consists of 19 pointings (Figure \ref{fig:FoV}).
The band 6 receivers were used in frequency range of 255--259, 271--275 GHz ($\sim$1.1 mm).
The data calibration are made using the Common Astronomy Software Application package (CASA; \citealt{2007ASPC..376..127M}).
We reconstruct a 1.1 mm continuum map with natural weighting. 
The synthesized beamsize and the rms level before the primary beam correction are 0\arcsec.53$\times$0\arcsec.41 and 55 $\mu$Jy/beam, respectively.

\begin{figure}
\begin{center}
\includegraphics[scale=1.0]{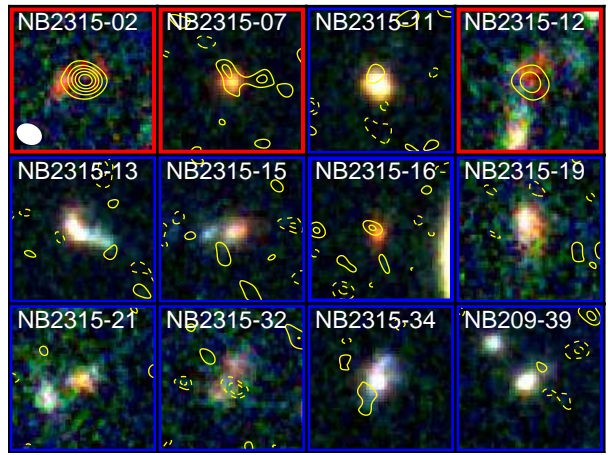}
\end{center}
\caption{1.1 mm continuum contours of our H$\alpha$-selected sample, superimposed on three-color images at ACS/$F814W$-, WFC3/$F125W$- and $F160W$-band (3\arcsec$\times$3\arcsec). 
Contours are plotted every 1$\sigma$, starting at 2$\sigma$, except for NB2315-02 and 12 (every 5$\sigma$).
Dashed contours denote negative fluxes.
The synthesized beamsize is shown at bottom left of the NB2315-02 panel.
\label{fig;hst_band6}}
\end{figure}

The SXDF-ALMA 1.5 arcmin$^2$ deep field covers 11 H$\alpha$-selected SFGs at $z=2.53$ and one at $z=2.19$ (Figure \ref{fig:FoV}).
As they are carefully identified on the basis of the narrow-band excess and the broad-band colors for the line separation between \ha and [O~{\sc iii}],
the redshift ranges are strongly-constrained by the width of narrow-band filters, $\Delta z=\pm0.02$.
X-ray detected AGNs or galaxies with a power-low spectral energy distribution (SED) at IRAC four bands are not included in our sample \citep{2013ApJ...778..114T}. 

\begin{figure*}[t]
\begin{center}
\includegraphics[scale=1.05]{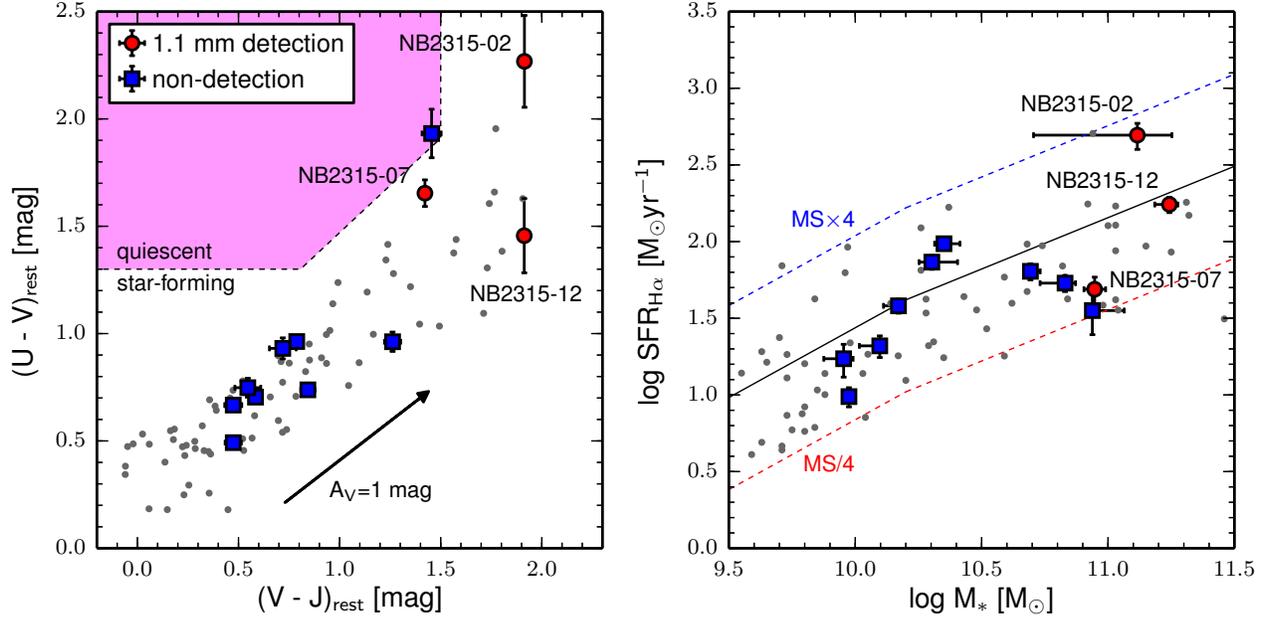}
\end{center}
\caption{
Left: Rest-frame $U-V$ versus $V-J$ color for our sample. Red circles and blue squares indicate 1.1 mm-detected and non-detection sources, respectively.
Gray circles show the parent sample of H$\alpha$-selected SFGs.
The red shaded area indicates a region where quiescent galaxies dominate \citep{2011ApJ...735...86W}.
Right: H$\alpha$-based star formation rates plotted against stellar masses. 
The black solid line indicates the main-sequence of SFGs at $z=2.0-2.5$ \citep{2014ApJ...795..104W}. 
\label{fig;UVJ}
}
\end{figure*}

First, we model their SEDs to measure the rest-frame $U-V$ and $V-J$ color ($UVJ$) using the 3D-HST catalog with photometries at 18 bands \citep{2014ApJS..214...24S} and the {\tt EAZY} code \citep{2008ApJ...686.1503B}.
Next, we estimate the stellar mass and the amount of dust extinction for our sample by fitting the photometric SEDs with the stellar population synthesis model of \citet{2003MNRAS.344.1000B}.
The SED fitting is done using the {\tt FAST} code \citep{2009ApJ...700..221K} with a solar metallicity, exponentially declining star formation histories (SFHs), and dust attenuation law of \citet{2000ApJ...533..682C}.
Following the recipe presented by \cite{2011ApJ...738..106W}, we adopt a minimum $e$-folding time of 300 Myr.
The uncertainties of the derived stellar mass are estimated from the 68\% confidence interval.
While the stellar mass tends to be the most robust parameter in SED modeling as long as single exponentially declining SFHs are assumed, the mass-to-light ratios could become smaller in 2-component SFHs with recent burst \citep{2007ApJ...655...51W}.
The differences are smaller for redder galaxies because a new burst makes galaxies blue in the rest frame optical. 
In section \ref{sec;gasmass}, the stellar masses are used for determine dust temperatures. 
A higher stellar mass leads to a lower dust temperature and results in a higher gas mass, but does not significantly change it for red galaxies.

\begin{table*}
\begin{center}
\caption{Properties of 1.1 mm-detected SFGs}{%
  \begin{tabular}{cccrrrrrr}
      \hline
      ID & R.A. & Decl. & $z_\mathrm{NB}$ & $M_*$ & $SFR_{\mathrm{H}\alpha}$ & $S_{1.1\mathrm{mm, aper}}$ & $M_\mathrm{gas}$ & $f_{\mathrm{gas}}$\\ 
       & (J2000) & (J2000) & & ($10^{10}~M_\odot$) & ($M_\odot$yr$^{-1}$) & (mJy) & ($10^{10}~M_\odot$) & (\%)\\ 
      \hline
      NB2315-02 & 02 17 40.53 & $-$05 13 10.7 & 2.53$\pm$0.02 & 13.1$^{+4.8}_{-8.0}$ &   495$\pm$95 & 2.06$\pm$0.13 & 10.3$^{+1.7}_{-1.5}$ & 44$^{+20}_{-8}$\\ 
      NB2315-07 & 02 17 42.67 & $-$05 13 58.4 & 2.53$\pm$0.02 & 8.9$^{+0.9}_{-0.8}$   &   49$\pm$10 & 0.38$\pm$0.15 & 2.7$^{+1.2}_{-1.0}$ &  23$^{+8}_{-8}$\\ 
      NB2315-12 & 02 17 41.11  & $-$05 13 15.2 & 2.53$\pm$0.02 & 17.5$^{+1.4}_{-2.2}$ &  174$\pm$20 & 1.34$\pm$0.15 & 8.1$^{+1.8}_{-1.3}$& 32$^{+5}_{-4}$\\ 
      \hline
    \end{tabular}}\label{tab:first}
\end{center}
\end{table*}

Star formation rates (SFRs) are derived from \ha luminosities \citep{1998ARA&A..36..189K} and SED-based extinctions, accounting for extra extinction toward H{\sc ii} regions \citep{2013ApJ...779..135W}.
The [N{\sc ii}] contribution has been corrected on the basis of the measured equivalent width of H$\alpha$+[N{\sc ii}] emission lines \citep{2013ApJ...778..114T}. 
We have also applied UNIMAP \citep{2015MNRAS.447.1471P} and extraction methods as described in \cite{2011A&A...532A..90L} to the archival Herschel-PACS data. 
We detect one source NB2315-02 at $S_{160\mu \mathrm{m}}=7.2\pm1.1$ mJy, corresponding to a total infrared luminosity of log $(L_\mathrm{IR}/L_\odot)=12.58$ using a conversion factor at $z=2.53$ from PACS 160 $\mu$m to $L_\mathrm{IR}$, based on the \cite{2011ApJ...738..106W} template.
The PACS 160 $\mu$m-based SFR is 414$\pm63$ $M_\odot$yr$^{-1}$ which is consistent with the H$\alpha$-based SFR of 495$\pm95$ $M_\odot$yr$^{-1}$.

\section{Results}
\label{sec;result}

\subsection{Detections in 1.1 mm continuum map}
\label{sec;dust_detection}

As the 1.1 mm continuum emission of SFGs at $z\sim2$ originates from thermal radiation of dust heated by massive stars,
it serves as a good tracer of dusty star-forming regions within galaxies. 
For all 12 SFGs in our sample, we measure the flux densities within the 1\arcsec.5 apertures and estimate the noise from 2000 random apertures ($1\sigma=125~\mu$Jy before the primary beam correction).
We adopt a $2\sigma$ detection criterion in the same manner as \cite{2014ApJ...783...84S}.
The probabilities where a negative 2$\sigma$ signal is detected by chance in the ALMA map is 2.2\%. 
Three SFGs are associated with a 1.1 mm source with a $>2.6\sigma$ significance (Figure \ref{fig;hst_band6}).
NB2315-02 and NB2315-12 are robustly detected above $9\sigma$, and NB2315-07 are marginally done at 2.6$\sigma$.
The spurious detection rates of 2.6$\sigma$ sources is 0.5\%.
Table \ref{tab:first} lists their measured flux densities along with stellar masses and SFRs.
The other 9 SFGs do not have a significant 1.1 mm continuum emission and their $2\sigma$ upper limit is $S_{1.1\mathrm{mm}}<290~\mu$Jy after the primary beam correction.

Our deep \ha narrowband selection covers wide ranges in colors and stellar mass of SFGs from blue, less massive to red, massive ones.
The 1.1 mm-detected SFGs are redder in the rest-frame optical and more massive than the non-detected sample (Figure \ref{fig;UVJ}).
The $UVJ$ diagram is useful for distinguishing between dusty star-forming, unobscured star-forming, and quiescent galaxies (e.g., \citealt{2007ApJ...655...51W}).
The three 1.1 mm sources are found to be dusty SFGs while most of SFGs without 1.1 mm detection fall in the unobscured star-forming regime, where also the bulk of galaxies satisfying the Lyman break criterion lies. 
One galaxy, which lies in the regime where quiescent galaxies dominate, is not detected above $2\sigma$. 
NB2315-07 is also close to this regime compared to NB2315-02 and NB2315-12.
Their lower position with respect to the main sequence supports that they are less star-forming galaxies.
We stack the unconfused 1.1 mm maps in the positions of 8 non-detected objects excluding one less star-forming galaxy to investigate the existence of faint emission for typical main-sequence SFGs with $\log(M_*/M_\odot)$=10.0-10.8.
However, the deep stacked image does not show a significant emission and gives the $2\sigma$ upper limit of $S_{1.1\mathrm{mm}}<100~\mu$Jy.

\subsection{Gas mass estimates}
\label{sec;gasmass}

Measurements of 1.1 mm continuum flux densities, $S_\nu$, allow us to derive gas masses of galaxies although there are uncertainties associated with dust temperature variations (e.g., \citealt{2014ApJ...783...84S,2015ApJ...800...20G}). 
We estimate the gas mass of our sample using a modified blackbody radiation model as, 

\begin{equation}
M_\mathrm{gas}=M_\mathrm{dust}/\delta_\mathrm{dgr}
=\frac{S_\nu d_\mathrm{L}^2}{\kappa_\mathrm{ISM} B_\nu (T_\mathrm{dust})(1+z)},
\end{equation}

\noindent
where $\delta_\mathrm{dgr}$ is dust-to-gas ratio, $\kappa_\mathrm{ISM}$ is the dust opacity per unit mass of interstellar medium ($\propto\nu^\beta$), $B_\nu$ is the Plank function, $T_\mathrm{dust}$ is the dust temperature, and $d_\mathrm{L}$ is the luminosity distance.
Here, $\kappa_\mathrm{ISM, 850\mu m}=4.84\times10^{-3}$ g$^{-1}$cm$^2$ and $\beta=1.8$ are adopted \citep{2014ApJ...783...84S}.
We note that for local galaxies the scatter of dust-to-gas ratio is $\pm$0.28 dex at fixed metallicities \citep{2014A&A...563A..31R}.
Although a conversion from monochromatic submillimeter luminosity to gas mass (i.e., $\kappa_\mathrm{ISM}$) is not straightforward, observations of dust emission still have a big cost advantage over CO observations. 
\citet{2015ApJ...800...20G} present scaling relations of dust temperature, gas fraction, and gas depletion timescale, by compiling CO and dust continuum data for each $\sim$500 galaxies over $0 < z < 3$, and find dust temperatures to only change slowly with specific SFR (sSFR) in the narrow range around the main sequence (see also \citealt{2014A&A...561A..86M}).
This scaling relation of dust temperatures is used for better estimates of the rest-frame 850 $\mu$m fluxes from the observed 1.1 mm fluxes.
The derived dust temperatures are 33$^{+3}_{-2}$, 28$\pm2$, and 30$\pm2$ K for NB2315-02, 07, and 12. 
The estimated gas masses and gas fractions, $f_\mathrm{gas}=M_\mathrm{gas}/(M_\mathrm{gas}+M_*)$, are summarized in Table \ref{tab:first}.

The 2$\sigma$ upper limit of our 1.1 mm survey corresponds to a gas mass of log $(M_{\rm gas}/M_\odot)=10.2$ in $T_\mathrm{dust}=30$ K at $z=2.53$.
Surprisingly, we do not detect the 1.1 mm emission from relatively less massive SFGs around the main-sequence although their gas mass is expected to be log $(M_{\rm gas}/M_\odot)=10.1-10.8$ based on the CO-based scaling relations of gas fraction with redshift, SFR, and stellar mass \citep{2015ApJ...800...20G}.
The non-detection could be partly caused by the usage of the fixed $\kappa_\mathrm{ISM}(=\kappa_\mathrm{dust}\delta_\mathrm{dgr})$ where the term of $\delta_\mathrm{dgr}$ is canceled out on the right side of the equation (1).
SFGs with lower metallicity should be fainter at 1.1 mm due to a lower dust-to-gas ratio than metal-rich ones, log $\delta_\mathrm{dgr}=-2+0.85\times$(12+log(O/H)-8.67), \citep{2011ApJ...737...12L} even if they have the same gas mass.
The mass-metallicity relation at $z\sim2$ predicts log $\delta_\mathrm{dgr}=-2.3$ for SFGs with log ($M_*/M_\odot$)=10 \citep{2014ApJ...789L..40W}.
Moreover, although the used dust opacity has been calibrated mainly by using outliers above the main-sequence such as local ultraluminous infrared galaxies and bright SMGs at $z\sim2$ \citep{2014ApJ...783...84S}, it is not necessarily appropriate for less massive normal SFGs at $z\sim2$.
Diffuse dust distributions within galaxies could lead to a low dust opacity \citep{2003Natur.424..285D}, resulting in faint 1.1 mm emission at fixed gas mass.
Also, dust SEDs of SFGs are better described by a multi-temperature model including cool dust in diffuse ISM (dominating a dust mass) and warm dust in birth clouds (dominating a total infrared luminosity).
Local normal SFGs have a cold component with $T_\mathrm{dust}\sim20$ K in a two-temperature model while a single-temperature model shows higher dust temperatures \citep{2011MNRAS.417.1510D}.

Assuming log $\delta_\mathrm{dgr}=-2.3$, $\kappa_\mathrm{dust, 850\mu m}=0.4$ g$^{-1}$cm$^2$ \citep{2003Natur.424..285D}, and $T_{\mathrm{dust}}=20$ K, 
the $2\sigma$ gas mass limit of our observations would become log $(M_{\rm gas}/M_\odot)=11.0$.
As the non-detections for less massive galaxies can be explained by a combination of these factors, it is difficult to determine whether the gas mass is actually as small as log $(M_{\rm gas}/M_\odot)=10.2$ with the 1.1 mm data alone.

\begin{figure}[t]
\begin{center}
\includegraphics[scale=1.00]{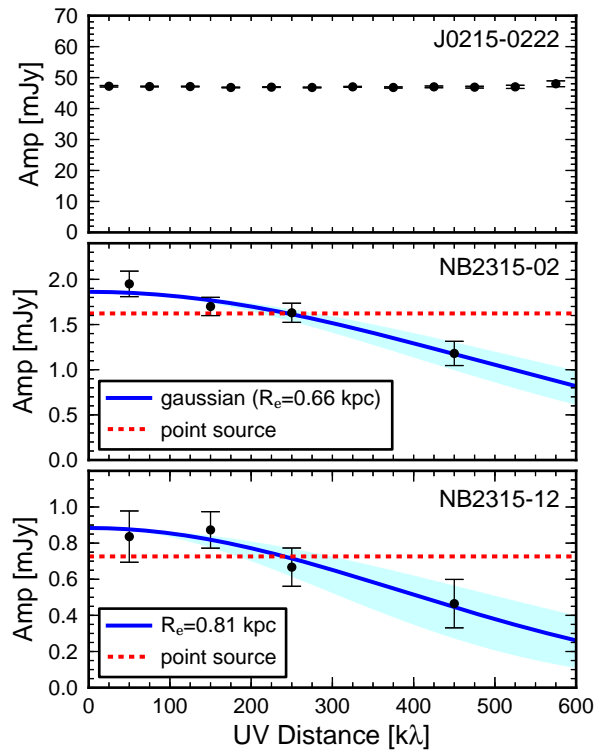}
\end{center}
\caption{
The visibility amplitudes averaged over $uv$ distances for the phase calibrator, J0215--0222 (top), NB2315-02 (middle), and NB2315-12 (bottom). 
A blue solid line and a shaded region indicate the best-fitting circular Gaussian component and the 1$\sigma$ error, respectively.
A red dashed line presents the best-fitting point source model as a reference.
\label{fig;uvfit}
}
\end{figure}

\subsection{Size measurements of dust emission}

We measure the size of the dust emission for two 1.1 mm bright galaxies (NB2315-02 and NB2315-12) by fitting the visibility data in the $uv$ plane.
Their signal-to-noise ratios in the 1.1 mm map are $>10$ in the flux density per synthesized beam to reliably constrain the size as the visibility coverage is similar as those by \citet{2014arXiv1411.5038I} and \citet{2015ApJ...799...81S}.
Then, we assume two models, a circular Gaussian component and a point source.
The {\tt uvamp} task in MIRIAD \citep{1995ASPC...77..433S} is used for calculating the visibility amplitudes averaged in annuli according to $uv$ distance after shifting the phase center to the center position measured in the image plane and subtracting a clean component of another source in the primary beam with the {\tt uvmodel} task.
The phase calibrator, J0215-0222, shows constant amplitudes as a function of $uv$ distance, suggesting a point source in the used antenna configuration (Figure \ref{fig;uvfit}).
On the other hand, two SFGs of our sample seem to be resolved at $>$300 k$\lambda$.
The gaussian fitting shows reduced chi-square values of 0.61 for NB2315-02 and 1.62 for NB2315-12 while the horizontal fitting (point source) does 5.76 and 2.60, respectively.
Therefore, we adopt the gaussian model which is the same approach as previous studies (\citealt{2014arXiv1411.5038I, 2015ApJ...799...81S}).
The best-fit results are FWHM=0\arcsec.16$^{+0.03}_{-0.02}$ ($R_e=0.66\pm^{+0.11}_{-0.08}$ kpc) for NB2315-02 and FWHM=0\arcsec.20$^{+0.06}_{-0.04}$ ($R_e=0.81^{+0.25}_{-0.15}$ kpc) for NB2315-12.
NB2315-02 surely has a compact, dusty star-forming component with $R_\mathrm{e}<$1 kpc at 3$\sigma$ significance. 
For a gaussian source with $R_e=0.66\pm^{+0.11}_{-0.08}$ kpc, 80$^{+7}_{-11}$\% (SFR$_{\mathrm{H}\alpha}$=396$^{+84}_{-93}~M_\odot$ yr$^{-1}$) of star formation traced by \ha is occurring within 1 kpc region.
Then, the SFR surface density could be 126$^{+27}_{-30}~M_\odot$ yr$^{-1}$kpc$^{-2}$. It is also $105^{+19}_{-22}~M_\odot$ yr$^{-1}$kpc$^{-2}$ in the case of usage of the PACS 160$\mu$m-based SFR.
The gas surface density is similarly estimated from the total gas mass to be $(2.6\pm0.5)\times10^{10}~M_\odot$ kpc$^{-2}$, which probably causes an extremely red color due to strong extinction (Figure \ref{fig;UVJ}).

\section{Discussion}

We find NB2315-02 to have a high SFR surface density, 126$^{+27}_{-30}~M_\odot$ yr$^{-1}$kpc$^{-2}$, corresponding to SFR=396 $^{+84}_{-93}~M_\odot$ yr$^{-1}$ within a region of 1 kpc radius. 
The central stellar surface density will become comparable with compact SFGs/QGs, $\log (M_*/R_e^{1.5})>10.3~[M_\odot$ kpc$^{-1.5}]$ \citep{2013ApJ...765..104B}, provided that the current star formation is maintained for another $50^{+12}_{-11}$ Myr in the galaxy center. 
This is plausible since it would need only 19$\pm3$\% of the total gas mass ($M_\mathrm{gas}=$10.3$^{+1.7}_{-1.5}\times10^{10}~M_\odot$) being converted into stars.
Then, the gas depletion timescale, $\tau_\mathrm{depl}\equiv M_\mathrm{gas}/SFR_{\mathrm{H}\alpha}$, is 207$^{+78}_{-44}$ Myr.
\cite{2014ApJ...795..145B} have also estimated dynamical masses of 13 compact SFGs from line widths of nebular emission (\ha or [O~{\sc iii}]) by near-infrared slit spectroscopy and derived similar gas depletion timescales ($\tau_\mathrm{depl}=$230$^{+110}_{-190}$ Myr).
The agreement of the gas depletion timescales would support that the compact dusty SFG can be an immediate progenitor of the high stellar surface density SFGs.


In the WFC3/$F160W$-band image, NB2315-02 has a sub-component with log $(M_*/M_\odot)=10.67$, which is seen 5 kpc east of the main component with the \ha peak (Figure \ref{fig;hst_band6}).
The compact starburst region appears to be located in between the two components.
Given the high gas surface density of $(2.6\pm0.5)\times10^{10}~M_\odot$ kpc$^{-2}$ in the compact dusty SFG, the rest optical morphology may be severely affected by strong attenuation.
One likely interpretation is that the two components would constitute a single large SFG with a central dusty star formation.
The gas fraction becomes $37^{+25}_{-3}$\% in this system including the companion. 
A physical process to reduce angular momentum is required in order to explain a concentrated gas distribution in a center of extended SFGs.
In gas-rich disks at $z\sim2$, gravitational torques and dynamical friction due to clumps can drive angular momentum out and cause mass inflow towards galaxy centers \citep{2014MNRAS.438.1870D, 2015MNRAS.450.2327Z}. 
Then, if the inflow timescale is shorter than the star formation timescale, a gas-rich, central starburst may form.
The measured high gas fraction of the compact dusty SFG could support the possibility of such a dissipational process. 
Another possible explanation is we are witnessing a late-stage merger with 1:3 stellar mass ratio immediately before the final coalescence.
Numerical simulations demonstrate that gas-rich mergers produce an instantaneous starburst at the final coalescence although it depends on merger parameters such as orbits of the two galaxies \citep{2013MNRAS.430.1901H}.

As compact SFGs are commonly defined by not high gas/dust surface densities but stellar ones \citep{2013ApJ...765..104B,2014ApJ...791...52B},
they are likely to have already completed most of morphological transformation from extended disks to compact spheroids while compact dusty SFGs have not done yet. 
NB2315-07 of our sample shows a compact morphology with a circularized, effective radius of $R_e=1.43\pm0.03$ kpc at $F160W$-band \citep{2014ApJ...788...28V} and satisfies the conditions for compact SFGs, $\log (M_*/R_e^{1.5})=10.71^{+0.05}_{-0.04}>10.3~[M_\odot$ kpc$^{-1.5}]$ and log ($SFR_{\mathrm{H}\alpha}/M_*)=-0.26\pm0.10>$ -1.0 [Gyr$^{-1}]$ \citep{2014ApJ...791...52B}.
The 1.1 mm-based gas fraction is 23$^{+8}_{-8}$\%, which is smaller than that of the compact dusty SFG at 2$\sigma$ significance.
Given the observed lower gas fraction and less star-forming properties shown in Figure \ref{fig;UVJ}, this compact less star-forming galaxy is likely to be already in a late stage of their evolutionary path to compact QGs.

In the fast quenching mode for galaxy evolution, we are looking a various compact objects in different evolutionary phases from dusty to quiescent through star-forming galaxies.
\cite{2014ApJ...791...52B} also present a diversity of compact SFGs from highly obscured (they are similar to compact dusty SFGs but already have a compact stellar core) to less star-forming ones (to which NB2315-07 belongs).
The spatial extent of gas remaining within galaxies can provide key information about the subsequent evolution of compact SFGs.
If gas is still concentrated in a galaxy center, compact SFGs are expect to exhaust all gas by a nuclear starburst or a feeding to a super massive black hole and then quench star formation.
Deep and high-resolution submillimeter imaging with ALMA has great potential to address this issue.

\


We thank the anonymous referee who gave us a number of comments, which improved the Letter.
This paper makes use of the following ALMA data: ADS/JAO.ALMA\#2012.1.00756.S. ALMA is a partnership of ESO (representing its member states), NSF (USA) and NINS (Japan), together with NRC (Canada), NSC and ASIAA (Taiwan), and KASI (Republic of Korea), in cooperation with the Republic of Chile. The Joint ALMA Observatory is operated by ESO, AUI/NRAO and NAOJ.
This work is based on observations taken by the 3D-HST Treasury Program (GO 12177 and 12328) with the NASA/ESA HST, which is operated by the Association of Universities for Research in Astronomy, Inc., under NASA contract NAS5--26555.
Data analysis were in part carried out on common use data analysis computer system at the Astronomy Data Center, ADC, of the National Astronomical Observatory of Japan.
This work was supported by JSPS KAKENHI Grant Numbers 21340045, 24244015, and 25247019.
RJI acknowledges support in the form of the ERC Advanced Grant, COSMICISM.




\end{document}